\documentstyle[aps,prl,epsfig]{revtex}
\draft
\newcommand{\size}{9.3}

\begin{document}
\title{Floquet scattering in parametric electron pumps}
\author{Sang Wook Kim}
\address{Max-Planck-Institut f{\"u}r Physik komplexer Systeme,
N{\"o}thnitzer Str. 38, D-01187 Dresden, Germany}
\date{\today}

\maketitle

\begin{abstract}

A Floquet scattering approach to parametric electron pumps is presented and compared with Brouwer's
adiabatic scattering approach [Phys. Rev. B {\bf 58}, R10135 (1998)] for a simple scattering 
model with two harmonically oscillating $\delta$-function barriers. For small strength of oscillating 
potentials these two approaches give exactly equivalent results while for large strength, 
these clearly deviate from each other. The validity of the adiabatic theory is also discussed by 
using the Wigner delay time obtained from the Floquet scattering matrix.

\end{abstract}



\section{introduction}

A parametric electron pump is a device that generates a dc current at zero bias potential through
cyclic change of system parameters. The most direct way to create a dc current was originally 
proposed by Thouless \cite{Thouless83}, who considered a system subjected to a traveling 
wave potential. This can be realized for example with the help of surface acoustic waves 
\cite{Talyanskii97}. Another possibility is to utilize quantum dots. In closed systems operating 
in the Coulomb blockade regime, integer number of electric charge can be transferred by sequential 
changes of barriers like a turnstile \cite{Kouwenhoven91}, whereas in open systems the electron 
pumping can be driven by adiabatic shape change in the confining potential or other parameters
which affect the interference pattern of the coherent electrons in the device. 
After a cycle of the adiabatic shape change we return to the initial configuration, but
the wavefunction may have its phase changed from the initial wavefunction. This is the
so-called geometric or Berry's phase \cite{Berry84}. This additional phase is equivalent to
some charges that pass through the quantum dot, namely, pumped charge \cite{Altshuler99,Avron00}.
Recently, adiabatic charge pumping in open quantum dots has attracted considerable attention
\cite{Altshuler99,Avron00,Brouwer98,Switkes99,Lubkin99,Zhou99,Entin-Wohlman02a,Entin-Wohlman02b}. 

Switkes et al. report an experimental investigation of electron pumping through an open quantum 
dot under the shape deformation controlled by two ac gate voltages \cite{Switkes99}. 
For weak pumping the dc voltage induced by the pumped dc current not only have a sinusoidal 
dependence on phase difference $\phi$ of the two ac voltages applied to the gates, but is also 
proportional to the square of the pumping strength $\lambda$. Many aspects of the experimental 
results can be understood in terms of Brouwer's scattering approach \cite{Brouwer98}. 
The change in the charge of the dot in response to a small variation of external parameter 
$\delta X$ is given by $\delta Q = e\sum \delta X_i dn/X_i$, where $dn/dX$ is the emissivity 
\cite{Buettiker94}. The pumped charge during each cycle can be determined by 
integrating $\delta Q$ along the closed path in the parameter space $\{X_i\}$ defined by one 
pumping cycle. Then, the pumped charge $Q$ is rewritten as an integral over the surface enclosed 
by the path by using Green's theorem, e.g. in two parameter case $Q=e\int_A dX_1 dX_2 \Pi(X_1,X_2)$.
Assuming $\Pi$ is constant over small area of $(X_1,X_2)$, for weak pumping one can show the 
pumping current $I_p \propto \omega \lambda^2 \sin \phi$, which exactly corresponds to what was 
observed in the experiment. For strong pumping, however, the dc voltage generated by the pumped 
current deviates from the sinusoidal dependence on $\phi$, and also departs from 
the $\lambda^2$ dependence. The non-sinusoidal dependence can be understood by taking into 
account that $\Pi$ is no longer constant over the integration area when $\lambda$ is considerably 
large. In Ref.~\cite{Switkes99} the anomalous $\lambda$ dependence was ascribed to the occurrence 
of significant heating and dephasing as a result of strong pumping.

The parametric electron pump is a time dependent system driven by (at least) two different time 
periodic perturbations with the same angular frequency and phase difference $\phi$. One can 
deal with this problem using not only adiabatic approximation exploited by Brouwer but also 
Floquet approach \cite{Sambe73}. An oscillating potential can transfer an incoming electron of 
energy $E$ to Floquet side bands at $E\pm n\hbar \omega$, where $n$ is an integer and $\omega$ 
is the angular frequency of the oscillation. A scattering matrix for a time dependent system
can be constructed from the interplay of these sidebands \cite{Li99,Henseler01}. So far there 
have been relatively few works on non-adiabatic parametric electron pumps \cite{Wagner00,Zhu02}, 
and the comparison between the adiabatic and the Floquet approach is still missing. We show, 
using a simple scattering model with two harmonically oscillating $\delta$-function barriers, 
that for small $\lambda$ these two approaches give exactly equivalent results for the pumped 
charge while for large $\lambda$ they are different. Even though the pumped 
current as a function of $\lambda$ obtained from these two approaches shows qualitatively 
similar behavior, the physical interpretation is completely different. This discrepancy in 
large $\lambda$ is not fully ascribed to the breakdown of adiabatic conditions since the Wigner 
delay times calculated by using Floquet scattering matrix are still much smaller than an 
oscillation period of external pumping potentials.

In Sec.~II, we introduce our model system and investigate the characteristics of the pumped current
using Brouwer's scattering approach. In Sec.~III, we study the Floquet approach for parametric 
electron pumps and compare with those of Sec.~II. We also discuss the validity of adiabatic theory 
by using the Wigner delay time. Finally, we conclude our paper in Sec.~IV.


\section{Brouwer's scattering matrix approach}

As a model system we consider 1D two harmonically oscillating $\delta$-function barriers with the
strengthes $X_1=V_1+\lambda_1 \cos \omega \tau$ and $X_2=V_2+\lambda_2 \cos (\omega \tau + \phi)$
respectively, separated by a distance $d$. This is a simplified model of the
experiment by Switkes et al., but possesses many important characteristics and can be easily
handled. Wei et al. studied parametric charge pumping aided by quantum resonance using this model
and found that the pumped current has large values near a resonance level \cite{Wei00}. Due to 
the double barrier geometry, resonant tunneling also plays an important role in charge pumping.

The $2 \times 2$ scattering matrix of the double barriers with the strengthes $X_1$ and $X_2$ is 
given by
\begin{equation}
S=\left( 
\begin{array}{cc}
r & t' \\
t & r'
\end{array}
\right),
\end{equation}
where $r$ and $t$ are the reflection and the transmission amplitudes respectively, for modes
incident from the left; $r'$ and $t'$ are similar quantities for modes incident from the right.
The charge emitted per cycle to the right is obtained from
\begin{equation}
Q_1 = e \int_0^{T_p} d\tau \left( \frac{dn_1}{dX_1}\frac{dX_1}{d\tau} 
+ \frac{dn_1}{dX_2}\frac{dX_2}{d\tau} \right),
\label{Brouwer_charge}
\end{equation}
where 
\begin{equation}
\frac{dn_1}{dX_m} = {\rm Im} 
\left( \frac{\partial t}{\partial X_m}t^* + \frac{\partial r'}{\partial X_m}r'^* \right)
\end{equation}
($m=1,2$), and $T_p (=2\pi/\omega)$ is the period of the pumping. One can show that the emitted 
charge to the left $Q_2$ is equal to $-Q_1$. Equation (\ref{Brouwer_charge}) can be rewritten 
in the following form by using Green's theorem
\begin{equation}
Q_1 = e \int_A dX_1 dX_2 \Pi (X_1,X_2),
\label{Brouwer_surface}
\end{equation}
where $\Pi (X_1,X_2) = \partial (dn_1/dX_2)/\partial X_1 - \partial (dn_1/dX_1)/\partial X_2$.
The pumped current is easily obtained from $I_1 = Q_1/T_p$. 

\begin{figure}
\center
\includegraphics[height=\size cm,angle=0]{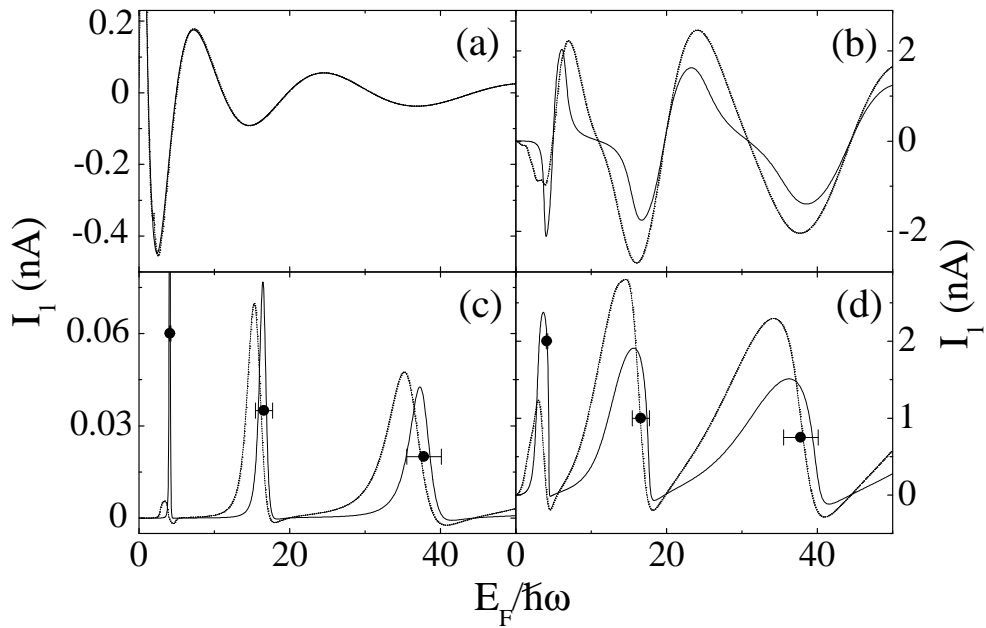}
\caption{The pumped current $I_1$ calculated by using Brouwer's approach (solid curves) 
and the Floquet approach (dotted curves) with $\phi=\pi/2$ for (a) $\lambda = 22.5$ 
meV$\cdot$nm and (b) $\lambda = 225$ meV$\cdot$nm with $V=0$, and (c) $\lambda = 22.5$ 
meV$\cdot$nm and (d) $\lambda = 225$ meV$\cdot$nm with $V=225$ meV$\cdot$nm. In (c) and (d) 
the transmission resonances are denoted by the filled circles ($\bullet$) obtained from 
considering static double barriers, whose $y$ values are chosen arbitrarily. The attached 
error bars represent the sizes of the imaginary energy of each resonance.}
\label{pcharge}
\end{figure}

We use the parameters $d=50$ nm for the distance between the two barriers, the effective mass 
$\mu = 0.067m_e$ of an electron in GaAs, and $T_p=9.09$ ps for the period of pumping, which 
corresponds to $\hbar \omega = 0.45$ meV. Figure \ref{pcharge} shows the pumped current as a 
function of the energy of an incident electron with $V_1 = V_2 = V$, $\lambda_1 = \lambda_2 = 
\lambda$, and $\phi=\pi/2$. We present two examples; the first is a nearly open case ($V$=0)
in Figs.~\ref{pcharge}a and \ref{pcharge}b, and the second is a closed case ($V$=225 meV$\cdot$nm)
in Figs.~\ref{pcharge}c and \ref{pcharge}d. All of them show interesting resonance-like structures.
In the closed case one can directly see the relation between the pumped currents and transmission 
resonances in Figs.~\ref{pcharge}c and \ref{pcharge}d, where the transmission poles are denoted by 
the filled circles ($\bullet$). To find these poles we considered a scattering problem with 
{\em static} double barriers of strength $V$. The imaginary part of the pole (denoted by the 
length of error bars attached to the filled circles) is related to the lifetime of a resonant 
(or quasibound) state, and determines the width of the transmission peaks. Peaks of the pumped 
current and their width are also described by the real and imaginary part of the resonance poles 
respectively, as shown in Fig.~\ref{pcharge}c. In Fig.~\ref{pcharge}d, however, the width of the 
pumped current is larger than the imaginary part of the resonance energy due to the effect of 
strong oscillation. Moreover, in the open case (Figs.~\ref{pcharge}a and \ref{pcharge}b), this 
analysis is no more relevant since there is no resonance in the static double barrier model.

\begin{figure}
\center
\includegraphics[height=\size cm,angle=0]{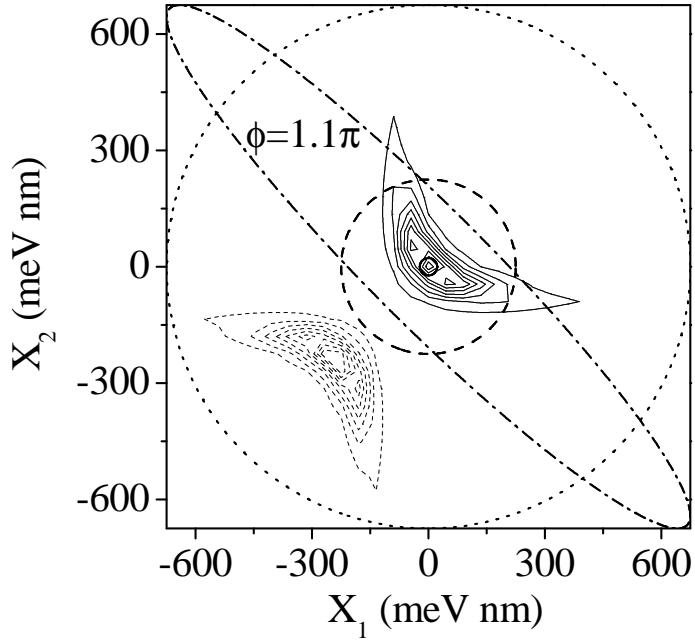}
\caption{Contour plot of flux $\Pi(X_1,X_2)$ for $E=6.005\hbar\omega$. 
The solid and the dashed contours represent positive and negative fluxes respectively,
in the range of $-0.016$ to $0.016$.
The circles represent the integration area for $\phi=\pi/2$ (see the text for detail) with 
$\lambda=22.5$ meV$\cdot$nm (thick solid curve), $\lambda=225$ meV$\cdot$nm (thick dashed curve), 
and $\lambda=675$ meV$\cdot$nm (thick dotted curve). The dashed dotted curve represents the 
integration area with $\phi=1.1\pi$ and $\lambda=675$ meV$\cdot$nm.}
\label{flux}
\end{figure}

Figure \ref{saturation} shows the pumped current as a function of the pumping strength $\lambda$
for $V=0$ (the open case) with $\phi=\pi/2$. Here for small $\lambda$ the pumped current depends 
on $\lambda^2$ and has an exact sinusoidal dependence on $\phi$, while for larger $\lambda$ the 
pumped current saturates and even decreases as $\lambda$ increases, and also deviates from the 
sinusoidal dependence on $\phi$. The decrease of the pumped current can be understood by 
considering the charge flux $\Pi(X_1,X_2)$ in Fig.~\ref{flux}. For small $\lambda$, the 
integration area only contains the positive $\Pi$'s (the solid circle). Hence, as $\lambda$ 
increases $\int_A \Pi$ also becomes larger, which is roughly proportional to the area enclosed 
by the circle, i.e. $\int_A \Pi \propto \lambda^2$. Around $\lambda=225$ meV$\cdot$nm (the dashed 
circle in Fig.~\ref{flux}) the integration area begins to include the region with both the positive 
and the negative $\Pi$'s, which means $\int_A \Pi$ no longer increases as $\lambda$ increases. 
This explains why the pumped current saturates and even decreases above $\lambda=225$ meV$\cdot$nm 
as shown in Fig.~\ref{saturation}. From this we expect that the behavior of the saturation of the 
pumped current would depend on $\phi$ since the shape of the integration area is determined by 
$\phi$. For example, with $\phi=1.1\pi$ the integration area denoted by the dashed dotted curve 
in Fig.~\ref{flux} is distorted from the original circular shape, so that even for $\lambda=675$ 
meV$\cdot$nm the integration area contains only positive $\Pi$'s. Figure \ref{saturation} clearly 
shows the pumped current with $\phi=1.1\pi$ saturates much more slowly in comparison with the pumped 
current with $\phi=\pi/2$. The non-sinusoidal dependence of the pumped current on $\phi$ for large 
$\lambda$ is also ascribed to the loss of circular symmetry of $\Pi$ in $(X_1,X_2)$. 


\section{Floquet approach}

Now we study the Floquet approach of the problem investigated in Sec.~II. Using the scattering 
matrix of a single $\delta$-function with sinusoidal time dependence (see Appendix for details),
we can obtain the total scattering matrix of the oscillating double $\delta$-functions in the 
following form 
\begin{equation}
S_{Fl}=\left( 
\begin{array}{cc}
{\bf r} & {\bf t'}\\
{\bf t} & {\bf r'} 
\end{array}
\right),
\end{equation}
where ${\bf r} = {\bf r}_L + {\bf t}_L({\bf I}-{\bf Qr}_R{\bf Qr}_L)^{-1}{\bf Qr}_R{\bf Qt}_L$, and 
${\bf t} = {\bf t}_R({\bf I}-{\bf Qr}_L{\bf Qr}_R)^{-1}{\bf Qt}_L$; ${\bf r}'$ and ${\bf t}'$ 
can also be obtained by replacing $L$ by $R$ in ${\bf r}$ and ${\bf t}$ respectively. Here 
${\bf r}_{R(L)}$ and ${\bf t}_{R(L)}$ are the reflection and the transmission matrices respectively, 
for the right (left) delta function with time dependence for modes from the left, and ${\bf I}$ is 
an identity matrix. Due to the reflection symmetry of each delta function 
${\bf r}'_{R(L)} = {\bf r}_{R(L)}$, and ${\bf t}'_{R(L)} = {\bf t}_{R(L)}$. During each one-way trip 
an electron at energy $E_n = E +n\hbar\omega$ picks up a phase factor $\exp(ik_nd)$, which is 
represented by the diagonal matrix ${\bf Q}_{mn}=\exp(ik_md)\delta_{mn}$. From $S_{Fl}$ we can 
obtain the total transmission coefficients for the propagating mode entering in the $m$th channel
\begin{equation}
T_{\rightarrow m}(E_{Fl})=\sum_{n=0}^{\infty}\left|{\bf t}_{nm}\right|^2,
\label{Fl_trans}
\end{equation}
where $E_{Fl}$ is the Floquet energy. The total transmission coefficient from the left to the right as 
a function of an energy of an incident electron is given by $T_\rightarrow(E)=T_{\rightarrow m}(E_{Fl})$ 
where $E=E_{Fl}+m\hbar\omega$. The total transmission from the right to the left $T_\leftarrow(E)$
can also be determined in a similar way. 

The pumped current to the right is given by \cite{Wagner00,Datta92}
\begin{equation}
I_1 = \frac{2e}{h} \int dE dE' [t(E',E) f_L(E) - t'(E,E') f_R(E')],
\label{datta_current}
\end{equation}
where $t(E',E)$ represents the transmission probability for scattering states incident from the
left at energy $E$ and emerging to the right at $E'$, and $t'(E,E')$ is defined in a similar
manner for the reverse direction. $f_L$ ($f_R$) is the Fermi-Dirac distribution in the left
(right) reservoir. Since the Floquet energy must be conserved during the scattering 
process, $E'$ and $E$ are given by $E_{Fl}+n\hbar\omega$ and $E_{Fl}+m\hbar\omega$ respectively 
($n$ and $m$ are integers). Using Eq.~(\ref{Fl_trans}), we can derive the relation 
$\int dE' t(E',E) = \sum_n {\bf t}_{nm}(E_{Fl}) = T_\rightarrow(E)$. 
Without external bias ($f_L = f_R = f$) Eq.~(\ref{datta_current}) can be rewritten as
\begin{equation}
I_1 = \frac{2e}{h} \int_0^\infty dE f(E) [T_\rightarrow(E)-T_\leftarrow(E)].
\label{current}
\end{equation}
At zero temperature it becomes
\begin{equation}
I_1 = \frac{2e}{h}\int_0^{E_F} dE [T_\rightarrow(E)-T_\leftarrow(E)],
\label{current_Fl}
\end{equation}
where $E_F$ is the Fermi energy. Equation (\ref{current_Fl}) can also be expressed as 
$I_1 = I_\rightarrow - I_\leftarrow$. This is quite interesting because from this point of view 
the pumped current merely corresponds to the difference of two currents having the opposite 
directions through a scatterer. 

\begin{figure}
\center
\includegraphics[height=\size cm,angle=0]{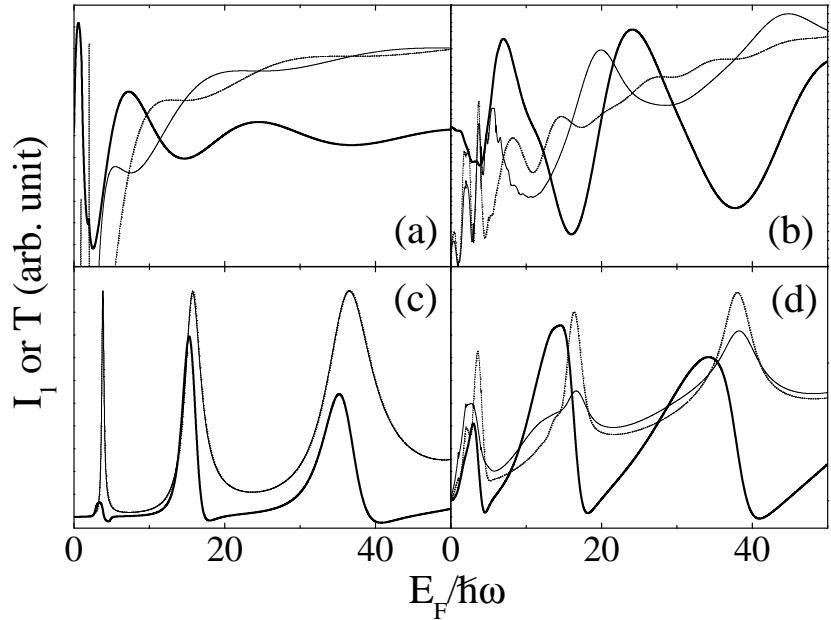}
\caption{Pumped current (thick solid curves), $T_\rightarrow$ (thin solid curves), 
and $T_\leftarrow$ (dotted curves) with the same parameters as used in Fig.~\ref{pcharge}. 
For clear comparison the unit and the scale of the y-axis are arbitrarily chosen.}
\label{t1t2}
\end{figure}

Figure \ref{pcharge} shows that in the open case ($V=0$) for small $\lambda$ (Fig.~\ref{pcharge}a)
the pumped current obtained from the Floquet approach is equivalent to that of Brouwer's while 
for rather larger $\lambda$ (Fig.~\ref{pcharge}b) they deviate from each other.
When $V \neq 0$ (the closed case), even for small $\lambda$ (Fig.~\ref{pcharge}c) they are 
quantitatively different near the resonances. Figure~\ref{t1t2} shows $T_\rightarrow$ and
$T_\leftarrow$ in comparison with the pumped current $I_1$. In the closed case 
(Figs. \ref{t1t2}c and \ref{t1t2}d) it is shown that $T_\rightarrow \approx T_\leftarrow$, 
and the transmission resonances are directly related to the maxima of the pumped currents. 
In the open case (Figs. \ref{t1t2}a and \ref{t1t2}b), however, 
$T_\rightarrow$ and $T_\leftarrow$ considerably differ from each other, and it is hard to 
determine the relation between the resonances and the pumped currents. In fact, $T_\rightarrow$
looks out of phase with $T_\leftarrow$, and their resonance structures are quite complicated. 
In the open case the resonance-assisted electron pumping \cite{Wei00} seems unclear even though 
the oscillatory behavior of transmission looks similar to that of the pumped current.

\begin{figure}
\center
\includegraphics[height=\size cm,angle=0]{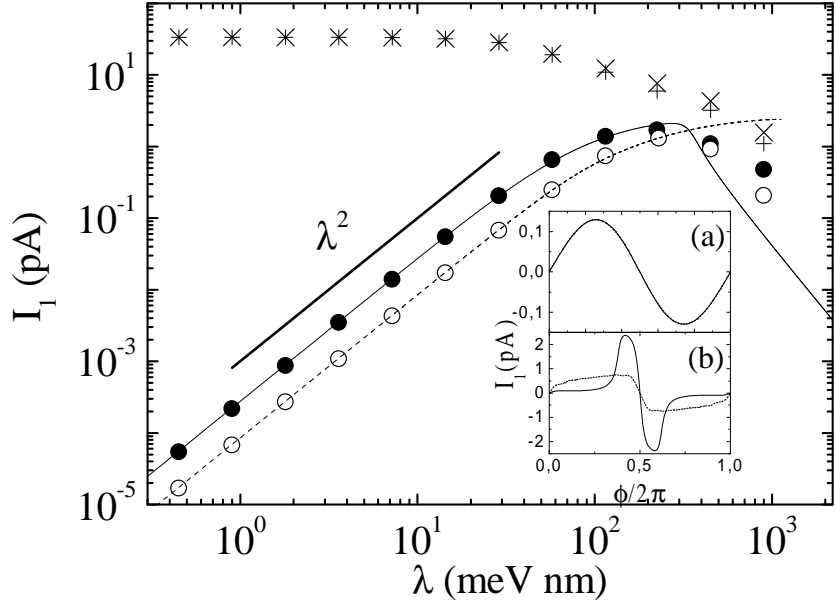}
\caption{The pumped current $I_1$ as a function of $\lambda$ with $V_1=V_2=0$ and 
$E=6.005\hbar\omega$ by using Brouwer's approach for $\phi=\pi/2$ (the solid curve) and 
$\phi=1.1\pi$ (the dashed curve), and the Floquet approach for $\phi=\pi/2$ ($\bullet$) and 
$\phi=1.1\pi$ ($\circ$). $\times$ and $+$ represent $I_\rightarrow$ and $I_\leftarrow$ for 
$\phi=\pi/2$ respectively. The inset shows the pumped current $I_1$ as a function of $\phi$ 
for (a) $\lambda = 22.5$ meV$\cdot$nm and (b) $\lambda = 675$ meV$\cdot$nm by using Brouwer's 
approach (the solid curves) and the Floquet approach (the dotted curves).}
\label{saturation}
\end{figure}

\begin{figure}
\center
\includegraphics[height=\size cm,angle=0]{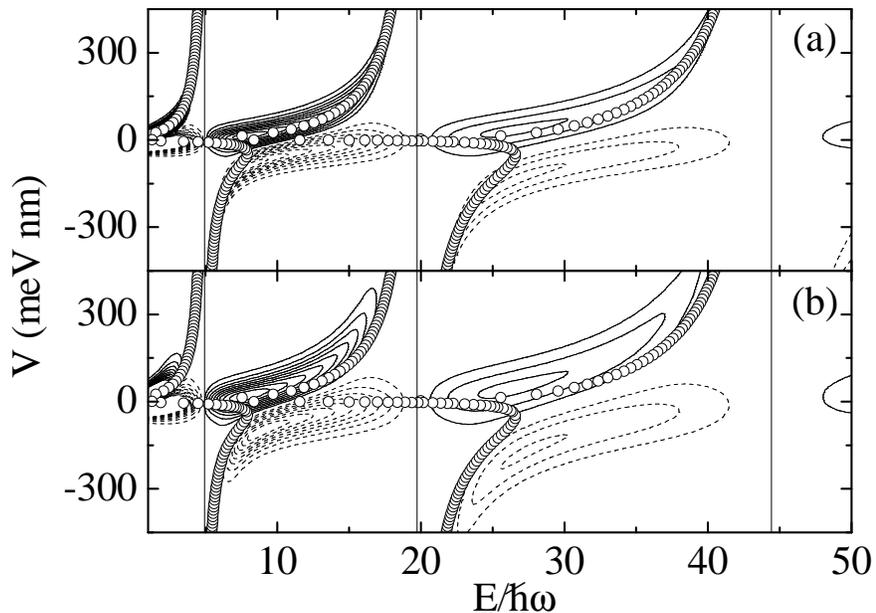}
\caption{Contour plot of the pumped current $I_1$ obtained from integrating $\Pi$ with $\lambda=22.5$ 
meV$\cdot$nm (see the text for detail) by using (a) Brouwer's approach and (b) the Floquet approach. 
The solid and the dashed contours represent positive and negative currents respectively, in the
range of $-0.2$ nA to $0.2$ nA.
The open circles ($\circ$) and the vertical lines represent 
the resonance energies of quasibound states of static double $\delta$-function barriers and the 
resonance energies from the condition $\pi^2\hbar^2m^2/2\mu d^2$ with an integer $m$ ($m=1,2,\cdots$)
respectively.} 
\label{comp}
\end{figure}

As $\lambda$ increases with $V=0$, in Fig.~\ref{saturation}, the pumped current obtained from the
Floquet approach also saturates and even decreases although their exact values are different 
from those of Brouwer's. We also plot $I_\rightarrow$ and $I_\leftarrow$ in Fig.~\ref{saturation}, 
where it is clear that in the Floquet approach the decrease of $I_1$ for large $\lambda$ is ascribed 
to the decrease of both $I_\rightarrow$ and $I_\leftarrow$. Usually the stronger the barrier strength,
the smaller the transmission. This interpretation differs from that of Brouwer's, in which 
the decrease of the pumped current is explained by considering the structure of $\Pi(X_1,X_2)$
(Sec. II). Since we cannot define $\Pi(X_1,X_2)$ explicitly in the Floquet formalism, we plot
$\Pi$ integrated over small circular area with a radius $\lambda$ centered at $(V_1,V_2)$.
We consider only 
the case that $V_1=V_2=V$. Figure \ref{comp} shows that the integrated $\Pi$'s obtained from the 
two different approaches for $\lambda=22.5$ meV$\cdot$nm look very similar to each other, and their 
overall structure are governed by the static double barrier resonances denoted by the circles ($\circ$). 

Brouwer's approach is based upon the adiabatic approximation, which implies that any time scale of 
the problem considered, especially the electron dwell time in a quantum dot (or inside double barriers 
in our case), must be much smaller than the period of the oscillation of a external pumping $T_p$
\cite{Brouwer98}. Using the Floquet formalism we can calculate the Wigner delay time $\tau_W$, 
which is the interaction time of the incident electron with the scattering potential \cite{Wigner55}
(see also \cite{Smith60,Fyodorov97}).
In this sense $\tau_W$ corresponds to the electron dwell time in the quantum pump. To obtain
the Wigner delay time we use the eigenvalues of the scattering matrix $S_{Fl}$. Due to the unitarity
of $S_{Fl}$ all the eigenvalues lie on the unit circle and can be written in the form 
$\exp(i\theta_\alpha)$. The Wigner delay time is defined by
\begin{equation}
\tau_W = \hbar \sum_\alpha \frac{d\theta_\alpha}{dE} \left|\left< k_n |\theta_{\alpha} \right>\right|^2,
\label{wigner_eq}
\end{equation}
where the eigenstate corresponding to the eigenvalue $\theta_{\alpha}$ and an input propagating 
state (or channel) with momentum $k_n$ are denoted by $\left| \theta_{\alpha} \right>$ and 
$\left| k_n \right>$ respectively \cite{Emmanouilidou01}. It is worth noting that the Wigner 
delay time is a function of the energy of the incident particle $E$ ($=E_{Fl}+n\hbar\omega$),
and $\left| k_n \right>$ and $\left| \theta_{\alpha} \right>$ are determined by $n$ and
$E_{Fl}$ respectively. If $\left< k_n |\theta_{\alpha} \right>$ is ignored in Eq.~(\ref{wigner_eq}) 
the Wigner delay time $\tau_W$ becomes trivial, i.e. $\tau_W(E+n\hbar\omega) = \tau_W(E)$.
Thus, we cannot observe any signature of the double barrier resonances. 

\begin{figure}
\center
\includegraphics[height=\size cm,angle=0]{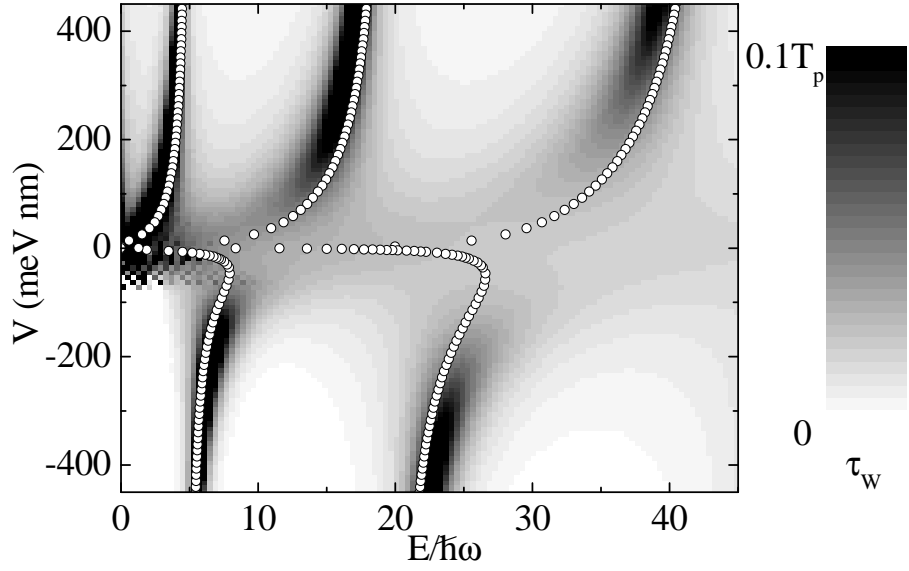}
\caption{Grey scale plot of the Wigner delay time. Black denotes times larger than 0.1$T_p$.  
The open circles ($\circ$) represent the resonance energies which are the same as in Fig.~\ref{comp}.}
\label{wigner}
\end{figure}

Figure \ref{wigner} shows the Wigner delay time using the same parameters exploited in Fig.~\ref{comp}.
The Wigner delay times become smaller for both larger $E$ and lower $V$, which can be understood 
when we take into account that usually the electron dwell time is short if the energy of an incident 
electron is large or the scattering barrier is weak. The Wigner delay times have larger values near the 
resonances, which is ascribed to the fact that at the resonances an electron can stay in the quantum dot 
for a long time. Near the resonances the adiabatic condition can break down.
This explains the deviation between the pumped currents observed in Figs.~\ref{pcharge}c.
In the open case ($V=0$) we also check the Wigner delay time for rather larger $\lambda$ 
(up to 675 meV$\cdot$nm), and observe they are smaller than $T_p$ by two orders of magnitude except for 
small incident energy $E$ (not shown), which means Brouwer's approach should be still applicable
even for large $\lambda$. This leads us to conclude that the deviations of the pumped current observed 
in Figs.~\ref{pcharge}a, \ref{pcharge}b and \ref{saturation} for large $\lambda$ are not simply ascribed 
to the breakdown of adiabatic condition. It is worth noting that recently it is shown that Brouwer's
formula is valid for arbitrary amplitudes of the modulating potential, as long as the lowest-order
adiabatic approximation can be employed \cite{Entin-Wohlman02a}. Let us just note that in 
Fig.~\ref{saturation} the discrepancy for large $\lambda$ appears when $|I_1|/|I_\rightarrow|$ 
(or $|I_1|/|I_\leftarrow|$) is not so small.

\begin{figure}
\center
\includegraphics[height=\size cm,angle=0]{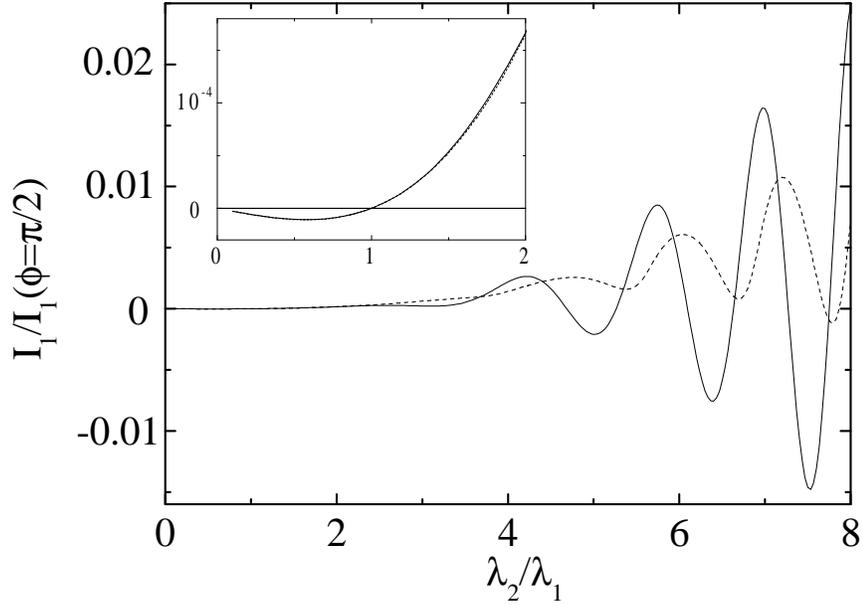}
\caption{Pumped current $I_1$ as a function of potential asymmetry $\lambda_2/\lambda_1$
with $\lambda_1=22.5$ meV$\cdot$nm and $E=6.005\hbar\omega$ at $\phi=0$ (solid curve) and
$\phi=\pi$ (dashed curve), where the currents are normalized to their values at at $\phi=\pi/2$.
Inset shows the magnification of a part of the plot.}
\label{asymmetry}
\end{figure}

One of the interesting consequences from Eq.~(\ref{current_Fl}) is that the pumped current still exists
even in the cases $\phi=0$ or $\pi$ when $\lambda_1 \neq \lambda_2$ \cite{Zhu02}. Since the integration 
area in parameter space is zero when $\phi=0$ or $\pi$, in Brouwer's approach the pumped current 
definitely vanishes. In contrast, even when $\phi=0$ or $\pi$, an asymmetry of the potential can 
lead to the asymmetry of the currents \cite{Datta92}, which is nothing but the pumped current in 
Eq.~(\ref{current_Fl}). Figure \ref{asymmetry} shows the pumed current as a function of ratio of the 
strength of two barriers $\lambda_2/\lambda_1$ with $\phi=0$ or $\pi$, and $E=6.005\hbar\omega$. 
Note that the pumped current is zero when $\lambda_1 = \lambda_2$. The oscillatory behavior is also 
related to the double barrier resonances.


\section{Summary}

We investigate the Floquet scattering in parametric electron pumps in comparison with Brouwer's adiabatic 
scattering approach exploiting two harmonically oscillating $\delta$-function barriers. In the Floquet
approach the pumped current simply corresponds to the difference of two currents having the opposite
direction through a scatterer. For small strength of the oscillating potentials, these two kinds of 
approach give exactly equivalent results while for large strength these show deviation. Even though
for large strength we obtain qualitatively similar results for the pumped currents using both approaches, 
the physical interpretation is completely different. The validity of the adiabatic approximation is also
discussed by calculating Wigner delay time. For large $\lambda$ although the adiabatic condition is 
well satisfied, a quantitative discrepancy between both approaches is still observed. In the Floquet 
approach, a non-zero pumped current can be obtained even when $\phi=0$ or $\pi$ (no current at all in 
Brouwer's approach), if the spatial symmetry of the potential is broken ($\lambda_1 \neq \lambda_2$).


\section*{Acknowledgments}
We would like to thank Henning Schomerus and Hwa-Kyun Park for helpful discussion.


\appendix
\section*{}

The scattering problem of a single $\delta$-function impurity with sinusoidal time dependence has been 
investigated by several authors \cite{Bagwell92,Martinez01,Kim02}. We would like to summarize how to 
construct its Floquet scattering matrix in this Appendix. The system is described by the Hamiltonian
\begin{equation}
H(x,t) = -\frac{\hbar^2}{2\mu}\frac{d^2}{dx^2} + [V_s + V_d \cos(\omega t + \phi)]\delta(x),
\end{equation}
where $\mu$ is the mass of the incident particle, while $V_s$ and $V_d$ represent the
strength of the static and the oscillating potential respectively. Using the Floquet
formalism the solution of this Hamiltonian can be expressed as
\begin{equation}
\Psi_{E_{Fl}}(x,t) = e^{-i E_{Fl} t/\hbar}\sum_{n=-\infty}^{\infty} \psi_n (x)
e^{-in\omega t},
\label{floquet}
\end{equation}
where $E_{Fl}$ is the Floquet energy which take continuous values in the interval
$0 < E_{Fl} \leq \hbar \omega$.

Since the potential is zero everywhere except at $x=0$, $\psi_n(x)$ is given by the
following form
\begin{equation}
\psi_n (x) = \left\{
\begin{array}{c}
A_n e^{ik_n x} + B_n e^{-ik_n x}, ~~~ x<0 \\
C_n e^{ik_n x} + D_n e^{-ik_n x}, ~~~ x>0,
\end{array} \right.
\label{plane_wave}
\end{equation}
where $k_n=\sqrt{2\mu(E_{Fl} + n\hbar\omega)}/\hbar$.
The wave function
$\Psi_{E_{Fl}}(x,t)$ is continuous at $x=0$,
\begin{equation}
A_n + B_n = C_n + D_n,
\label{eq1.1}
\label{osc_bc1}
\end{equation}
and the derivative jumps by
\begin{equation}
\left.\frac{d\Psi_{E_{Fl}}}{dx}\right|_{x=0^+} - \left.\frac{d\Psi_{E_{Fl}}}{dx}\right|_{x=0^-}
=\frac{2m}{\hbar^2}[V_s + V_d \cos(\omega t + \phi)]\Psi_{E_{Fl}}(0,t).
\label{osc_bc2}
\end{equation}
Using Eq.~(\ref{floquet}) this leads to the condition
\begin{eqnarray}
&& ik_n (C_n - D_n - A_n + B_n) \nonumber \\
&=& \gamma_s(A_n+B_n) +\gamma_d(e^{-i\phi}A_{n+1}+e^{i\phi}A_{n-1}+e^{-i\phi}B_{n+1}+e^{i\phi}B_{n-1}) \label{eq2} \\
&=& \gamma_s(C_n+D_n) +\gamma_d(e^{-i\phi}C_{n+1}+e^{i\phi}C_{n-1}+e^{-i\phi}D_{n+1}+e^{i\phi}D_{n-1}) \nonumber
,\end{eqnarray}
where $\gamma_s=2\mu V_s/\hbar^2$ and $\gamma_d=\mu V_d/\hbar^2$. After some algebra
we have the following equation from Eqs.~(\ref{eq1.1}) and (\ref{eq2})
\begin{equation}
\left(
        \begin{array}{c}
        \vec{B} \\ \vec{C}
        \end{array}
\right)
=
\left(
        \begin{array}{cc}
        -(I+\Gamma)^{-1}\Gamma & (I+\Gamma)^{-1} \\
        (I+\Gamma)^{-1} & -(I+\Gamma)^{-1}\Gamma
        \end{array}
\right)
\left(
        \begin{array}{c}
        \vec{A} \\ \vec{D}
        \end{array}
\right)
\label{mat_eq}
,\end{equation}
where
\begin{equation}
\Gamma =
\left(
        \begin{array}{ccccc}
        \ddots & \ddots & 0 & 0 & 0 \\
        \gamma_d e^{i\phi}/ik_{-1} & \gamma_s/ik_{-1} & \gamma_d e^{-i\phi}/ik_{-1} & 0 & 0 \\
        0 & \gamma_d e^{i\phi}/ik_0 & \gamma_s/ik_0 & \gamma_d e^{-i\phi}/ik_0 & 0 \\
        0 & 0 & \gamma_d e^{i\phi}/ik_1 & \gamma_s/ik_1 & \gamma_d e^{-i\phi}/ik_1\\
        0 & 0 & 0 & \ddots & \ddots
        \end{array}
\right),
\end{equation}
and $I$ is an infinite-dimensional square identity matrix. Eq.~(\ref{mat_eq}) can also be
expressed in the
form $\left.|{\rm out}\right> = M \left.|{\rm in}\right>$, where $M$ connects the input
coefficients to the output coefficients including the associated evanescent Floquet
sidebands. In order to construct the scattering matrix we multiply an identity to both sides,
$K^{-1}K\left.|{\rm out}\right> = M K^{-1}K\left.|{\rm in}\right>$, where $K_{nm}=\sqrt{k_n}
\delta_{nm}$. Then we have $\vec{J}_{out}=\bar{M}\vec{J}_{in}$, where $\vec{J}$ represents
the amplitude of probability flux and $\bar{M} \equiv KMK^{-1}$. It should be mentioned
that $\bar{M}$ is not unitary due to the evanescent modes included.

If we keep only the propagating modes, we obtain the unitary scattering matrix $S$
\cite{Li99,Henseler01}, which can be expressed in the following form
\begin{equation}
S = \left(
        \begin{array}{cccccc}
        r_{00} & r_{01} & \cdots & t'_{00} & t'_{01} & \cdots \\
        r_{10} & r_{11} & \cdots & t'_{10} & t'_{11} & \cdots \\
        \vdots & \vdots & \ddots & \vdots  & \vdots  & \ddots \\
        t_{00} & t_{01} & \cdots & r'_{00} & r'_{01} & \cdots \\
        t_{10} & t_{11} & \cdots & r'_{10} & r'_{11} & \cdots \\
        \vdots & \vdots & \ddots & \vdots  & \vdots  & \ddots \\
        \end{array}
\right),
\label{osc_smatrix}
\end{equation}
where $r_{nm}$ and $t_{nm}$ are the reflection and the transmission amplitudes respectively,
for modes incident from the left; $r'_{nm}$ and $t'_{nm}$ are similar quantities for modes
incident from the right.


\bibliographystyle{prsty}


\newpage

\end{document}